\documentclass{appolb}
\usepackage[]{graphicx}
\usepackage{amsmath}
\usepackage{amsfonts}
\usepackage{amssymb}
\usepackage{graphicx}
\usepackage[toc,page]{appendix}%
\begin{document}
\title{Strong decays of excited vector mesons.%
\thanks{Presented at Excited QCD 2017, 7-13 May 2017, Sintra, Portugal.}%
}
\author{M. Piotrowska$^{1}$, F. Giacosa$^{1,2}$
\address{$^1$ Institute of Physics, Jan Kochanowski University,\\\textit{ul. Swietokrzyska 15, 25-406, Kielce, Poland. }\\
$^{2}$ Institute for Theoretical Physics, J. W. Goethe University,\\\textit{ Max-von-Laue-Str. 1, 60438 Frankfurt, Germany.}}
\\
}
\maketitle
\begin{abstract}
We study two nonets of excited vector mesons, predominantly corresponding to radially excited vector mesons with quantum numbers $n\hspace{0.08cm} ^{2S+1}L_{J}$\\$=2^3S_1$ and to orbitally excited vector mesons with quantum numbers $n\hspace{0.08cm} ^{2S+1}L_{J}=1^3D_1$. We evaluate two types of decays of these mesons: into two pseudoscalar mesons and into a pseudoscalar and a ground-state vector meson. We compare the results with experimental data taken from PDG. We also make predictions for the strange-antistrange state in the $1^3D_1$ nonet denoted as $\phi(1930)$, which has not yet been discovered. 
\end{abstract}
  
\section{Introduction}
Quantum Chromodynamics (QCD) is the theory describing the strong interactions
between quarks and gluons, which build up the hadrons. Mesons, in general, are
bosonic hadrons, while a so-called \textquotedblleft conventional meson" is a
quark-antiquark ($q\bar{q}$) state \cite{godfrey}. For the discussion of \textquotedblleft non-conventional" mesons which are
predominantly four-quark or gluonic states see Refs.
\cite{klempt,pelaezrev,xyzrev,ochs}.\newline In the low-energy sector (i.e., with
light quarks $u,$ $d,$ and $s$), numerous experiments and theoretical
calculations have provided a wide range of information about $q\bar{q}$
ground-state mesons with various quantum numbers \cite{pdg,qqpdg}. On the
contrary, excited states are rather poorly determined. Yet, the study of
excited states is necessary to confirm the validity of the $q\bar{q}$ picture
and to identify states beyond the conventional assignment. Quite
interestingly, two nonets of excited vector mesons were experimentally
identified: \{$\rho(1450)$, $K^{\ast}(1410)$, $\omega(1420)$, $\phi(1680)$\}
and \{$\rho(1700)$, $K^{\ast}(1680)$, $\omega(1650)$, $\phi(???)$\}. They
predominately correspond to radially excited, $n\hspace{0.08cm}^{2S+1}%
L_{J}=2^{3}S_{1},$ and orbitally excited, $1^{3}D_{1},$ spectroscopic
configurations (both of them with $J^{PC}=1^{--}$ as the standard,
ground-state vector mesons). Even if experimental data require improvement,
there is enough information on masses and decays for a systematic analysis of
these resonances. A summary about both nonets is presented in Tab. 1. \\
In this work, following the forthcoming publication \cite{exvm}, we use a QFT
model based on flavor symmetry in order to study the strong decays of the two
nonets mentioned above. This type of approach has been already used in the
past in a series of papers on tensor mesons \cite{tensor}, pseudovector mesons
\cite{pseudovector}, pseudotensor mesons \cite{pseudotensor}, as well as the
more controversial scalar mesons \cite{gutsche}. Hence, it appears naturally and
promising to perform this study for excited vector mesons as well. As an
outcome of our approach, we shall show that the interpretation of the mesons
listed above as standard $q \bar{q}$ states is upheld. Moreover, we can make
predictions for the decay widths of the not-yet measured orbitally excited
$s\bar{s}$ state, denoted as $\phi(???)\equiv$ $\phi(1930),$
where $1930$ MeV is an estimate of its mass from the relation $m_{\phi
(???)}-m_{\phi(1680)}\simeq$ $m_{\rho(1700)}-m_{\rho(1450)}\simeq250$ MeV
\cite{exvm}.
 
\begin{table}
\renewcommand{\arraystretch}{1.25}
\begin{tabular}[c]{|c|c|c|}
\hline Type of excitation & Radially excited  & Angular momentum excited \\ &vector mesons & vector mesons\\
\hline
Quantum numbers & $n\hspace{0.08cm} ^{2S+1}L_{J}=2^3S_1$&$n\hspace{0.08cm} ^{2S+1}L_{J}=1^3D_1$\\
\hline
Notation&$V_E$& $V_D$\\
\hline
S&$1 \uparrow\uparrow $ & $1\uparrow\uparrow$\\
\hline
n&2&1\\
\hline
L&0&2\\
\hline
Orbital & 
\includegraphics[width=0.08 \textwidth] {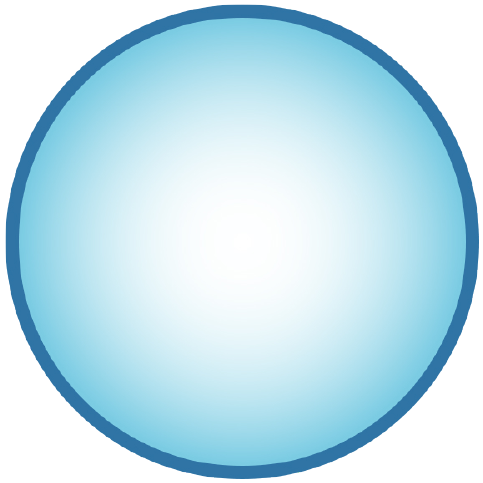}

& \includegraphics[width=0.1 \textwidth] {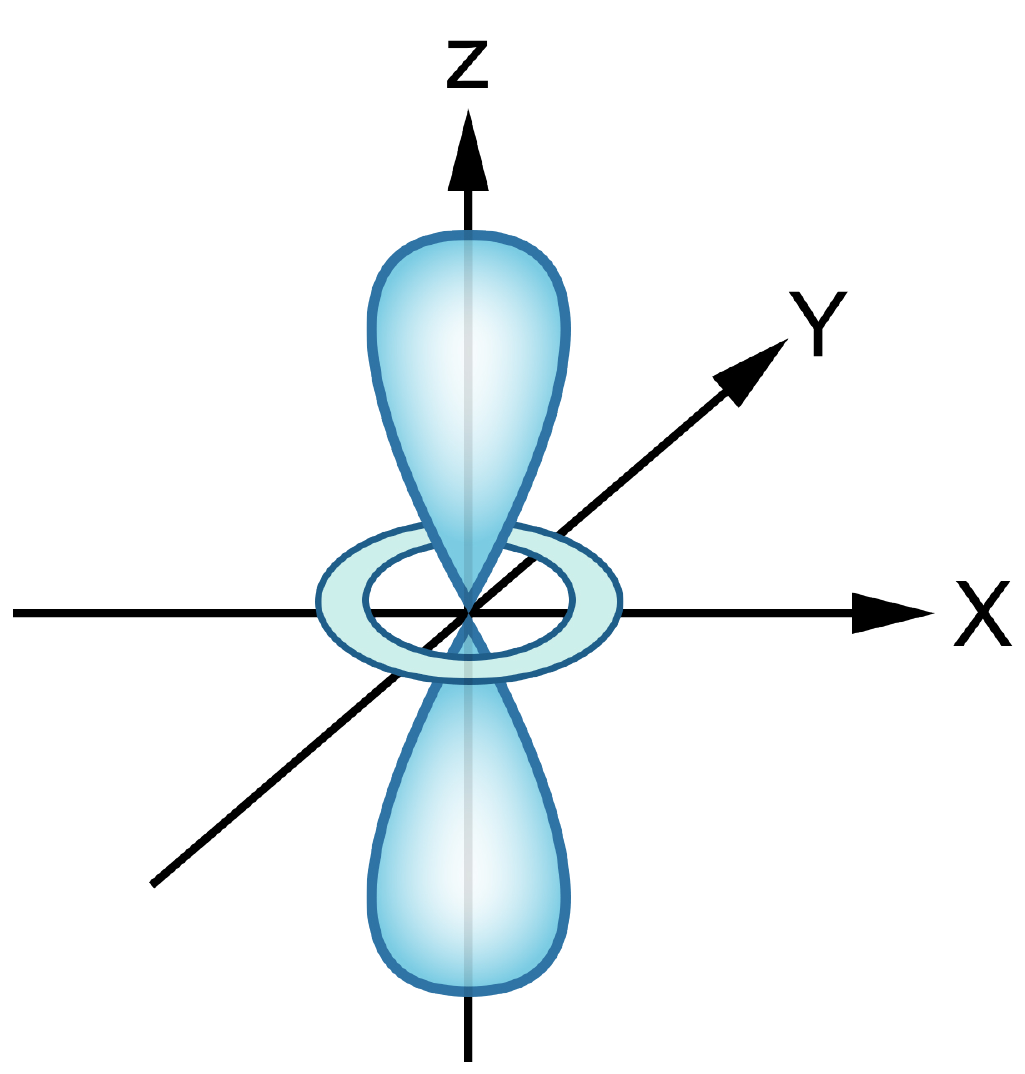}\\
\hline
Radial function &\includegraphics[width=0.25 \textwidth] {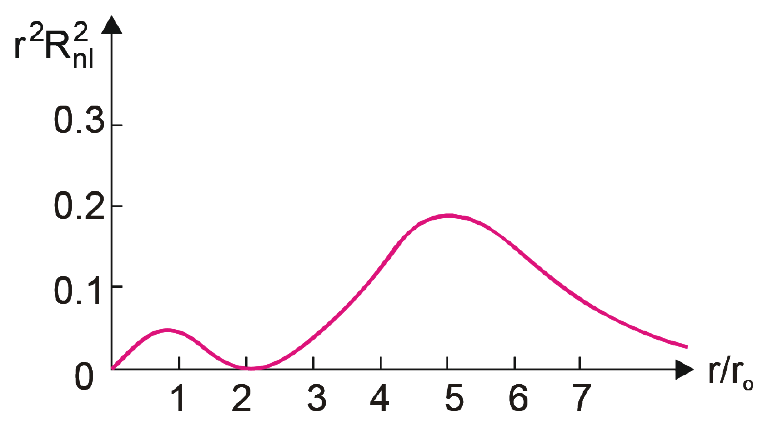}&\includegraphics[width=0.25 \textwidth] {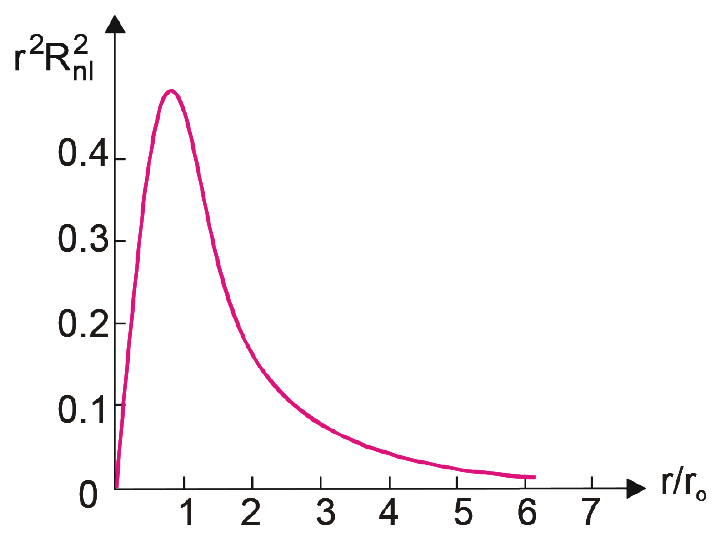}\\
\hline
Associated states&$ \rho (1450), K^* (1410)$,&$ \rho (1700), K^* (1680)$,\\
 &$\omega(1420), \phi(1680)$& $\omega(1650), \phi(1930)$\\
 \hline
Decay types& $V_E\rightarrow PP$& $V_D \rightarrow PP$\\
&$V_E \rightarrow VP$&$V_D \rightarrow VP$\\
\hline
\end{tabular}
\caption{Basic informations about radially and orbitally excited vector mesons.}
\end{table}

\section{The model}
In our approach, we use a relativistic QFT model that couples four nonets of
mesons. It is invariant under parity $P$, charge conjugation $C$, and
$U(3)_{V}$ flavour transformation. The Lagrangian of our model reads

\begin{align}
\nonumber
\mathcal{L}=ig_{EPP}Tr\left(  [\partial^{\mu}P,V_{E,\mu}]P\right)
+ig_{DPP}Tr\left(  [\partial^{\mu}P,V_{D,\mu}]P\right)   \\ 
+  g_{EVP}Tr\left(
\tilde{V}_{E}^{\mu\nu}\{V_{\mu\nu},P\}\right)  +g_{DVP}Tr\left(  \tilde{V}%
_{D}^{\mu\nu}\{V_{\mu\nu},P\}\right).  \label{lag}%
\end{align}
Here, $P$ represents the matrix of pseudoscalar mesons \{$\pi,K,\eta
(547),\eta^{\prime}(958)$\}, $V_{\mu}$ the matrix of ground-state vector
mesons \{$\rho(770),$ $K^{\ast}(892),$ $\omega(782)$, $\phi(1020)$\}, while 
$V_{E,\mu}$ contains
\{$\rho(1450)$, $K^{\ast}(1410)$, $\omega(1420)$, $\phi(1680)$\} and
$V_{D,\mu}$ contains \{$\rho(1700)$, $K^{\ast}(1680)$, $\omega(1650)$,
$\phi(???)$\}. The Lagrangian describes decays of the type $V_{E(D)}%
\rightarrow PP$ (first two terms) and of the type $V_{E(D)}\rightarrow VP$ (third and fourth terms).
Moreover,  $[.,.]$ is the usual commutator, $\{.,.\}$ the anticommutator, and
$\tilde{V}_{E(D)}^{\mu\nu}=\frac{1}{2}\epsilon^{\mu\nu\alpha\beta}%
(\partial_{\alpha}V_{E(D),\beta}-\partial_{\beta}V_{E(D),\alpha})$ are the
dual fields. According to our model, the tree-level decay widths of a resonance
$R$ ($R=V_{E}\hspace{0.2cm}\text{or}\hspace{0.2cm}R=V_{D}$) read:
\begin{equation}
\Gamma_{R\rightarrow PP}=\frac{s_{RPP}|\vec{k}|^{3}}{6\pi m_{R}^{2}}\left(
\frac{g_{RPP}}{2}\lambda_{RPP}\right)  ^{2}\text{, }\Gamma_{R\rightarrow
VP}=\frac{s_{RVP}|\vec{k}|^{3}}{12\pi}\left(  \frac{g_{RVP}}{2}\lambda
_{RVP}\right)  ^{2}\text{,}%
\end{equation}
where $\vec{k}$ is the tree-momentum of an outgoing resonance (see, e.g., the
review on kinematics in the PDG \cite{pdg}), $s_{RPP}$ an isospin
factor, and $\lambda_{RPP}$ a coefficient which is determined by expanding the
traces in Eq. (\ref{lag}) \cite{exvm}. Note, Eq. (\ref{lag}) implements only flavour symmetry,
hence there are no chiral partners. An extension of the present approach to
full chiral symmetry is possible following Refs. \cite{dick,vg}; this is left
for the future. 
\section{Results}
The Lagrangian (\ref{lag}) contains four coupling constants $g_{EPP}$%
, $g_{EVP}$, $g_{DPP}$, $g_{DVP}$. In order to determine them, we use
experimental data from PDG: $\Gamma_{K^{\ast}(1410)\rightarrow K\pi},$ the
total width of $\phi(1680),$ the ratio $\frac{\Gamma_{K^{\ast}(1680)\rightarrow K\rho}}{\Gamma_{K^{\ast}(1680)\rightarrow K\pi}}$ %
, and also
$\Gamma_{K^{\ast}(1680)\rightarrow
K\pi},$ out of which we obtain: $g_{EPP}=3.66\pm0.4$, $g_{EVP}=18.4\pm3.8,$
$g_{DPP}=7.15\pm0.94,$ and $g_{DVP}=16.5\pm3.5$.

The results for the decays into two pseudoscalar mesons and into a
pseudoscalar and a ground-state vector meson are presented in Tab. 2, where a
comparison to some experimental data listed in PDG is shown. 
\begin{table}[h!]
\renewcommand{\arraystretch}{1.25}
\begin{tabular}[c]{|c||c|c||c|c|}
\hline
&\multicolumn{2}{|c||}{{\bf Radially excited}} & \multicolumn{2}{|c|}{{\bf Orbitally excited}}\\
&\multicolumn{2}{|c||}{{\bf vector mesons [MeV]}} & \multicolumn{2}{|c|}{{\bf vector mesons [MeV]}}\\
\hline
\hline
{\bf Decay channel}&{\bf Theory}& {\bf Exp.}& {\bf Theory}& {\bf Exp.}\\
\hline
\multicolumn{5}{|c|}{${\bf V_R \rightarrow PP}$}\\
\hline
\hline
$\rho_{R}\rightarrow\bar{K}K$ & $6.6\pm1.4$ & $<$ 6.7 $\pm$1.0& $40\pm11$ & $8.3^{+10}_{-8.3}$\\
\hline
$\rho_{R}\rightarrow\pi\pi$ & $30.8\pm6.7$ & $\sim$ 27 $\pm$ 4&$140\pm37$ & $75\pm30$\\
\hline
$K^{\ast}_{R}\rightarrow K\pi$ & $15.3\pm3.3$ & 15.3 $\pm$ 3.3&$82\pm22$ & $125\pm43$ \\
\hline
$K^{\ast}_{R}\rightarrow K\eta$ & $6.9\pm1.5$ & -&$52\pm14$ & -\\
\hline
$K^{\ast}_{R}\rightarrow K\eta^{\prime}$ & $\approx0$ & -& $0.72\pm0.02$ & -\\
\hline
$\omega_{R}\rightarrow\bar{K}K$ & $5.9\pm1.3$ & -&$37\pm10$ & - \\
\hline
$\phi_{R}\rightarrow\bar{K}K$ & $19.8\pm4.3$ & +&$104\pm28$ & * \\
\hline
\multicolumn{5}{|c|}{${\bf V_R \rightarrow VP}$}\\
\hline
\hline
$\rho_{R}\rightarrow\omega\pi$ & $74.7\pm31.0$ & $\sim84\pm13$& $140\pm59$ & + \\
\hline
$\rho_{R}\rightarrow K^{\ast}(892)K$ & $6.7\pm2.8$ & + & $56\pm23$ & $83 \pm 66$\\
\hline
$\rho_{R}\rightarrow\rho(770)\eta$ & $9.3\pm3.9$ & $<16.0\pm2.4$& $41\pm17$ & $68 \pm 42$ \\
\hline
$\rho_{R}\rightarrow\rho(770)\eta^{\prime}$ & $\approx0$ & -& $\approx0$ & -\\
\hline
$K^{\ast}_{R}\rightarrow K\rho$ & $12.0\pm5.0$ & $<16.2\pm1.5$& $64\pm27$ & $101\pm35$\\
\hline
$K^{\ast}_{R}\rightarrow K\phi$ & $\approx0$ & -& $13\pm6$ & -\\
\hline
$K^{\ast}_{R}\rightarrow K\omega$ & $3.7\pm1.5$ & -& $21\pm9$ & -\\
\hline
$K^{\ast}_{R}\rightarrow K^{\ast}(892)\pi$ & $28.8\pm12.0$ & $>93\pm8$& $81\pm34$ & $96\pm33$\\
\hline
$K^{\ast}_{R}\rightarrow K^{\ast}(892)\eta$ & $\approx0$ & -& $0.5\pm0.2$ & -\\
\hline
$K^{\ast}_{R}\rightarrow K^{\ast}(892)\eta^{\prime}$ & $\approx0$ & -& $\approx0$ & -\\
\hline
$\omega_{R}\rightarrow\rho\pi$ & $196\pm81$ & dominant& $370\pm156$ & $205\pm23$ \\
\hline
$\omega_{R}\rightarrow K^{\ast}(892)K$ & $2.3\pm1.0$ & -& $42\pm18$ & -\\
\hline
$\omega_{R}\rightarrow\omega(782)\eta$ & $4.9\pm2.0$ & -& $32\pm13$ & $56 \pm 30$\\
\hline
$\omega_{R}\rightarrow\omega(782)\eta^{\prime}$ & $\approx$ 0 & -& $\approx0$ & -\\
\hline
$\phi_{R}\rightarrow K\bar{K}^{\ast}$ & 110 $\pm$ 46 & dominant& $260\pm109$ & *\\
\hline
$\phi_{R}\rightarrow\phi(1020)\eta$ & 12.2 $\pm$ 5.1 & +& $67\pm28$ & *\\
\hline
$\phi_{R}\rightarrow\phi(1020)\eta^{\prime}$ & $\approx$ 0 & -& $\approx0$ & *\\
\hline
\end{tabular}
\caption{Values of decay widths for two types of decay of excited vector mesons.  (+) stands that the decay channel was seen, while (-) was not seen and (*) means that decaying resonance has not yet been discovered.}
\end{table}
We use the following notation: $(\rho_{R},K_{R}^{\ast},\omega_{R},\phi_{R})$ is either
identified with \{$\rho(1450),$ $K^{\ast}(1410),$ $\omega(1420),$ $\phi
(1680)$\} or with  \{$\rho(1700),$ $K^{\ast}(1680),$ $\omega(1650),$
$\phi(1930)$\}, respectively. 
The decay channels which are large according to
theory are also large in experiments. Vice-versa, theoretically small decays
were not measured yet, hence our results are predictions.  In general, there
is a good agreement of our theory with data, hence the interpretation of the
two nonets as radially excited and orbitally excited vector mesons is
confirmed (see, however, the discussions in\ Ref. \cite{coito}). For a more
detailed description of various decay ratios and some open issue  concerning
conflicting experimental results, we refer to \cite{exvm}. Surely, new
experimental data would be welcome.  For the putative state
$\phi(1930)$ there are four different decay channels. Even if this resonance
is expected to be broad, it could be measured in the future at the GlueX
\cite{gluex} and CLAS12 \cite{clas12} experiments at Jefferson lab.

\section{Conclusions}
In this work, we have reported on strong decays of the (predominantly) radially
excited vector mesons \{$\rho(1450)$, $K^{\ast}(1410)$, $\omega(1420)$,
$\phi(1680)$\} and the (predominantly) orbitally excited vector mesons
\{$\rho(1700)$, $K^{\ast}(1680)$, $\omega(1650)$, $\phi(???)\equiv\phi
(1930)$\} (see Tab. 1 for a summary) by using an effective relativistic QFT
Lagrangian based on flavour symmetry \cite{exvm}. In total, we evaluated 48
different decay channels for both nonets of mesons and compared them to known
experimental values, see Tab. 2. The overall agreement confirms the validity
of the $q\bar{q}$ assignment for these states. Radiative decays can be also studied
without no new parameters, see Ref. \cite{exvm}. Moreover,  we made predictions for a
not yet discovered resonance  $\phi(???)\equiv\phi(1930)$ belonging to the
heavier, predominantly $1^{3}D_{1}$ nonet. We expect new experimental
data from GlueX and CLAS12 experiments to compare it to our results. \newline

\textbf{Acknowledgements: } The authors thank C. Reisinger for cooperation and S. Coito, D. Parganlija, V. Sauli, and J. Sammet for discussions. The authors acknowledge support from the Polish National Science Centre (NCN) through the OPUS project no. 2015/17/B/ST2/01625.



\end{document}